\documentclass[aps,english,twocolumn,showkeys,prb,preprintnumbers,superscriptaddress,floatfix]{revtex4-1}
\usepackage{amsmath,amssymb}
\usepackage{color}
\usepackage{bm}
\usepackage{textcomp} % This package is just to give the text quote '
\usepackage{enumitem}
\usepackage{hyperref}
\usepackage{natbib}
\usepackage{color,soul}
\usepackage{graphicx}
\usepackage{pgffor}
\usepackage[final]{pdfpages}
\makeatletter
\AtBeginDocument{\let\LS@rot\@undefined}
\makeatother
\allowdisplaybreaks
\begin{document}
	
	\title{Halide perovskites under polarized light: Vibrational symmetry analysis using polarized
		Raman}
	\author{Rituraj Sharma}
	\thanks{These authors contributed equally}
	\affiliation{Department of Materials and Interfaces, Weizmann Institute of Science, Rehovoth 76100, Israel}
	\author{Matan Menahem}
	\thanks{These authors contributed equally}
	\affiliation{Department of Materials and Interfaces, Weizmann Institute of Science, Rehovoth 76100, Israel}
	\author{Zhenbang Dai}
	%\thanks{These authors contributed equally}
	\affiliation{Department of Chemistry, University of Pennsylvania, Philadelphia, Pennsylvania 19104, USA}
	\author{Lingyuan Gao}
	\affiliation{Department of Chemistry, University of Pennsylvania, Philadelphia, Pennsylvania 19104, USA}
	\author{Thomas M. Brenner}
	\affiliation{Department of Materials and Interfaces, Weizmann Institute of Science, Rehovoth 76100, Israel}
	\author{Lena Yadgarov}
	\affiliation{Department of Materials and Interfaces, Weizmann Institute of Science, Rehovoth 76100, Israel}
	\author{Jiahao Zhang}
	\affiliation{Department of Chemistry, University of Pennsylvania, Philadelphia, Pennsylvania 19104, USA}
	\author{Yevgeny Rakita}
	\affiliation{Department of Materials and Interfaces, Weizmann Institute of Science, Rehovoth 76100, Israel}
	\author{Roman Korobko}
	\affiliation{Department of Materials and Interfaces, Weizmann Institute of Science, Rehovoth 76100, Israel}
	\author{Iddo Pinkas}
	\affiliation{Department of Materials and Interfaces, Weizmann Institute of Science, Rehovoth 76100, Israel}
	\author{Andrew M. Rappe}
	\email{rappe@sas.upenn.edu} 
	\affiliation{Department of Chemistry, University of Pennsylvania, Philadelphia, Pennsylvania 19104, USA}
	\author{Omer Yaffe}
	\email{omer.yaffe@weizmann.ac.il}
	\affiliation{Department of Materials and Interfaces, Weizmann Institute of Science, Rehovoth 76100, Israel}
	\date{\today}
	
	\begin{abstract}
		\noindent In the last decade, hybrid organic-inorganic halide perovskites have emerged as a new type of semiconductor for photovoltaics and other optoelectronic applications. Unlike standard, tetrahedrally bonded semiconductors (e.g. Si and GaAs), the ionic thermal fluctuations in the halide perovskites (i.e. structural dynamics) are strongly coupled to the electronic dynamics. Therefore, it is crucial to obtain accurate and detailed knowledge about the nature of the atomic motions within the crystal. This has proved to be challenging due to low thermal stability and the complex, temperature-dependent structural phase sequence of the halide perovskites. Here, these challenges are overcome and a detailed analysis of the mode symmetries is provided in the low-temperature orthorhombic phase of methylammonium-lead iodide. Raman measurements using linearly- and circularly- polarized light at 1.16 eV excitation are combined with density functional perturbation theory (DFPT). By performing an iterative analysis of Raman polarization-orientation dependence and DFPT mode analysis, the crystal orientation is determined. Subsequently, accounting for birefringence effects detected using circularly polarized light excitation, the symmetries of all the observed Raman-active modes at 10 K are assigned. 
	\end{abstract}
	
	\keywords{Halide Perovskites, Methylammonium-lead iodide, Polarization-orientation Raman, orthorhombic, birefringence}
	
	\maketitle
	
	\newpage
	
	\section{Introduction}
	Methylammonium lead iodide (MAPI) has emerged as an attractive material for next-generation photovoltaic technologies, owing to its optimal bandgap and excellent photoconversion efficiency\cite{Brenner2016,Gratzel2014,Stranks2015,Stoumpos2013,Wehrenfennig2014}. From a fundamental standpoint, MAPI and other hybrid organic-inorganic halide perovskites (HOPs) are intriguing because unlike standard semiconductors, the optical and charge transport properties of HOPs are considered to be strongly coupled to their lattice fluctuations\cite{Panzer2017,Guo2017,Egger2018,Mayers2018,Whalley2016b}.
	
	Indeed, there has been intensive research dedicated to deciphering the structural characteristics of halide perovskites, employing X-ray diffraction (XRD), Raman and neutron scattering, dielectric measurements, \textit{etc}\cite{Baikie2015,Swainson2015,Rahman2020,Sendner2016,Anusca2017}. Specifically, Raman scattering is a highly adaptable technique that gives qualitative and quantitative insights into local structure and dynamics. There are many recent reports discussing the Raman spectra of HOPs\cite{Yaffe2017a,Guo2017,Perez-Osorio2018}. Density functional theory (DFT) and molecular dynamics (MD) based calculations have been used to interpret the Raman spectra and find the energies of the Raman-active modes\cite{Brivio2015a,Quarti2014}. However, Raman measurements on MAPI have proven to be very challenging both from the experimental and computational standpoints. Regarding the experiment, the high light absorption coefficient and low thermal stability of MAPI do not allow long exposure to intense visible light without inducing irreversible sample damage\cite{Park2015,Pistor2016,Ruan2019}. This is true even for near-IR, 1.58 eV light\cite{Ledinsky2015a}. Furthermore, the complex crystal structure and temperature phase sequence do not permit an unambiguous determination of the crystallographic orientation at low temperatures without a separate XRD measurement in identical conditions. Regarding DFPT computation, the complex crystal structure and the large number of atoms in the unit cell result in a very large number of Raman-active modes that are closely spaced in energy. Therefore, without prior knowledge of the mode symmetry, it is very difficult to determine which specific modes are observed in the experiment.
	
	Here, we overcome the aforementioned challenges by performing an iterative analysis between  polarization-orientation (PO) Raman scattering measurements with 1.16 eV laser excitation and DFPT computation. We provide an unambiguous determination of the Raman tensors for all the observed Raman-active modes as well as the crystallographic orientation in MAPI at 10 K. Furthermore, by combining our PO measurements with circularly polarized light excitation, we are able to account for the effect of birefringence on the Raman signal. Thus, we demonstrate the great potential of our measurements and analysis as an exceptional tool for vibrational mode assignments in highly complicated systems like HOPs.
	
	\section{Results and Discussion}
	MAPI has an \textit{ABX${}_{3}$} perovskite structure, with PbI${}_{6}$ in corner-sharing octahedra and MA${}^{+}$ inside the cuboctahedral cavities. It is known to have three structural phases: an orthorhombic (\textit{Pnma}) phase below 162 K, a tetragonal (I4/mcm) phase between 162 and 327 K and a cubic (\textit{Pm-3m}) structure  beyond 327 K\cite{Poglitsch1987,Whitfield2016}.
	
\begin{figure}[ht!]
		\includegraphics[width=3.5in, keepaspectratio=true]{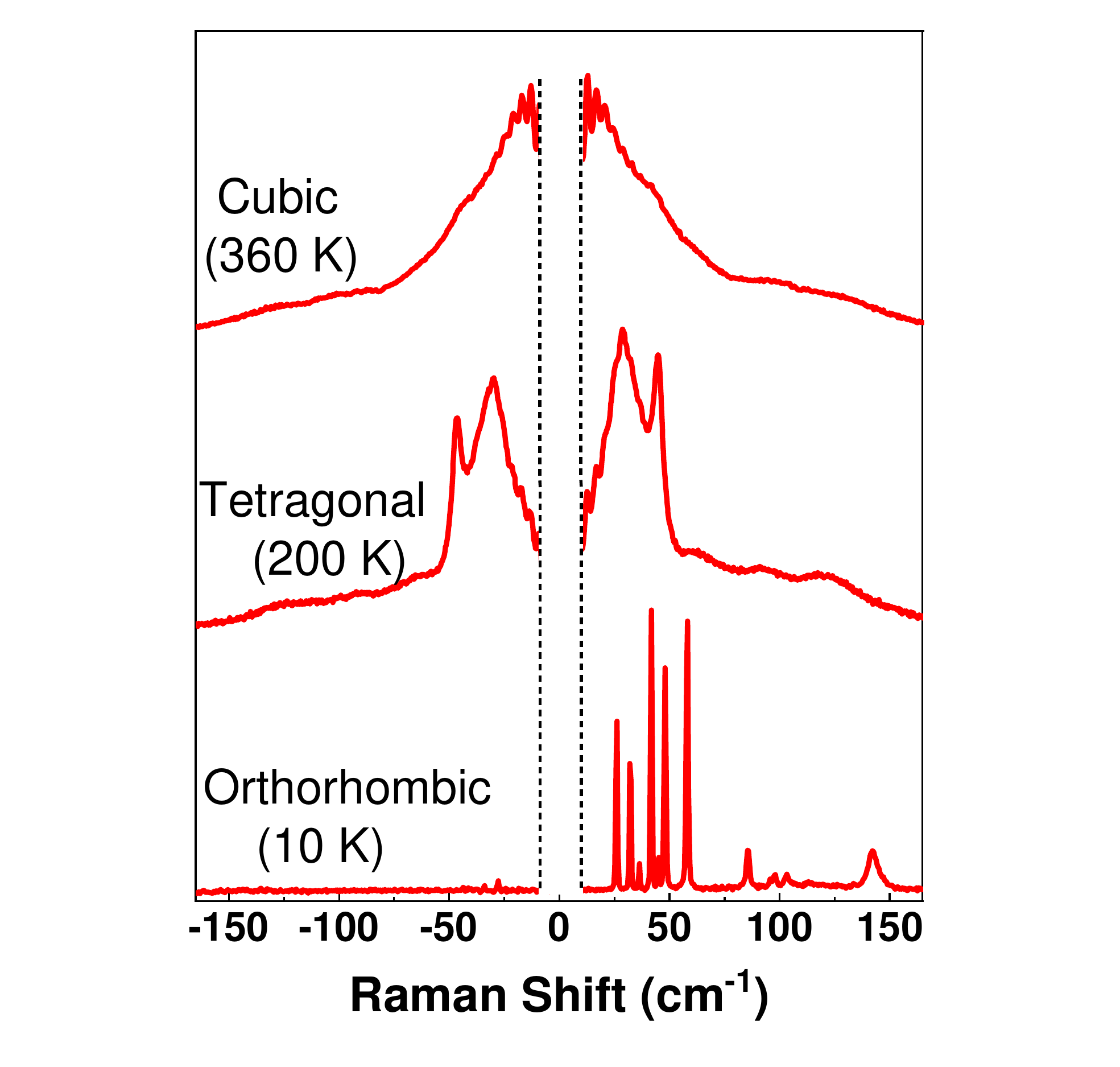}
		\caption{Temperature dependence of unpolarized Raman spectra showing the phase transitions in MAPI. The Rayleigh scattering region of the spectra is removed from 0 cm$^{-1}$ to the vertical dotted lines.}
		\label{Ramanphases}
	\end{figure}

	Fig.~\ref{Ramanphases} shows the unpolarized Raman spectra of MAPI single crystals taken across the phase sequence covering simultaneously both Stokes and anti-Stokes scattering. The spectrum of the orthorhombic phase (10 K) shows sharp peaks, and the orthorhombic-tetragonal phase transition is characterized by an abrupt change, where the sharp peaks are replaced by relatively broad features.
	
	The Raman spectra in the tetragonal and cubic phases are envelopes of the orthorhombic phase spectrum. Therefore, to interpret the nature of atomic motions in the high-temperature phases, a thorough understanding of the vibrational characteristics of the low-temperature phase is a prerequisite. In this paper, we focus on the Raman spectra of the orthorhombic phase acquired at 10 K.
	
	\subsection{PO Raman scattering analysis}\label{PO}
	The atomic vibrations in a crystal are closely related to the crystallographic arrangement of atoms and have specific symmetries. Group theory provides tools to predict the symmetries of the vibrational normal modes by decomposing the crystal space group to its irreducible representations. 
	According to factor group analysis, we expect 18 Raman-active lattice modes in the orthorhombic MAPI structure with the following irreducible representation\cite{Kroumova2003,Schuck2018}:
	
	\begin{equation} 
	\mathit{\Gamma}={5A}_g+{4B}_{1g} +{5B}_{2g}+{4B}_{3g} 
	\end{equation}
	
	The Raman tensor of each mode, which relates the scattering cross section to light polarization vector, must obey the symmetry of the irreducible representation as well\cite{Hayes2004,Wilson1980}. The corresponding Raman tensors are given as\\ 
	
	\noindent $A_g=\left( \begin{array}{ccc}
	a & 0 & 0 \\ 
	0 & b & 0 \\ 
	0 & 0 & c \end{array}
	\right)$, $B_{1g}=\left( \begin{array}{ccc}
	0 & d & 0 \\ 
	d & 0 & 0 \\ 
	0 & 0 & 0 \end{array}
	\right)$,  $B_{2g}=\left( \begin{array}{ccc}
	0 & 0 & e \\ 
	0 & 0 & 0 \\ 
	e & 0 & 0 \end{array}
	\right)$ and $B_{3g}=\left( \begin{array}{ccc}
	0 & 0 & 0 \\ 
	0 & 0 & f \\ 
	0 & f & 0 \end{array}
	\right)$.\\
	
	To extract the Raman tensor for each mode, we perform PO Raman measurements. Briefly, the crystal is excited by plane-polarized laser light
	having polarization e$ _{i} $. The scattered light e$ _{s} $ is then filtered
	by another polarizer (analyzer) for polarizations parallel
	and perpendicular to the incident light. This measurement is repeated after rotating the polarization of the
	incident light. The Raman tensor for each peak can be identified from the angular dependence of the peak intensity according to\cite{Kranert2016}:
	
	\begin{equation}  
	I\left(\theta \right)\propto {\left|e_s(\theta ){\mathcal{R}}^TR\mathcal{R}e_i(\theta )\right|}^2 
	\label{Placzek}
	\end{equation} 
	
	\noindent where $e_i$ and $e_s$ are the light polarization vectors which are either parallel or perpendicular to one another, and $\mathcal{R}$ is a rotation matrix transforming the Raman tensor \textit{R} from the crystal system to the laboratory system. 
	
	\begin{figure*}[ht!]
		\includegraphics[width=6in]{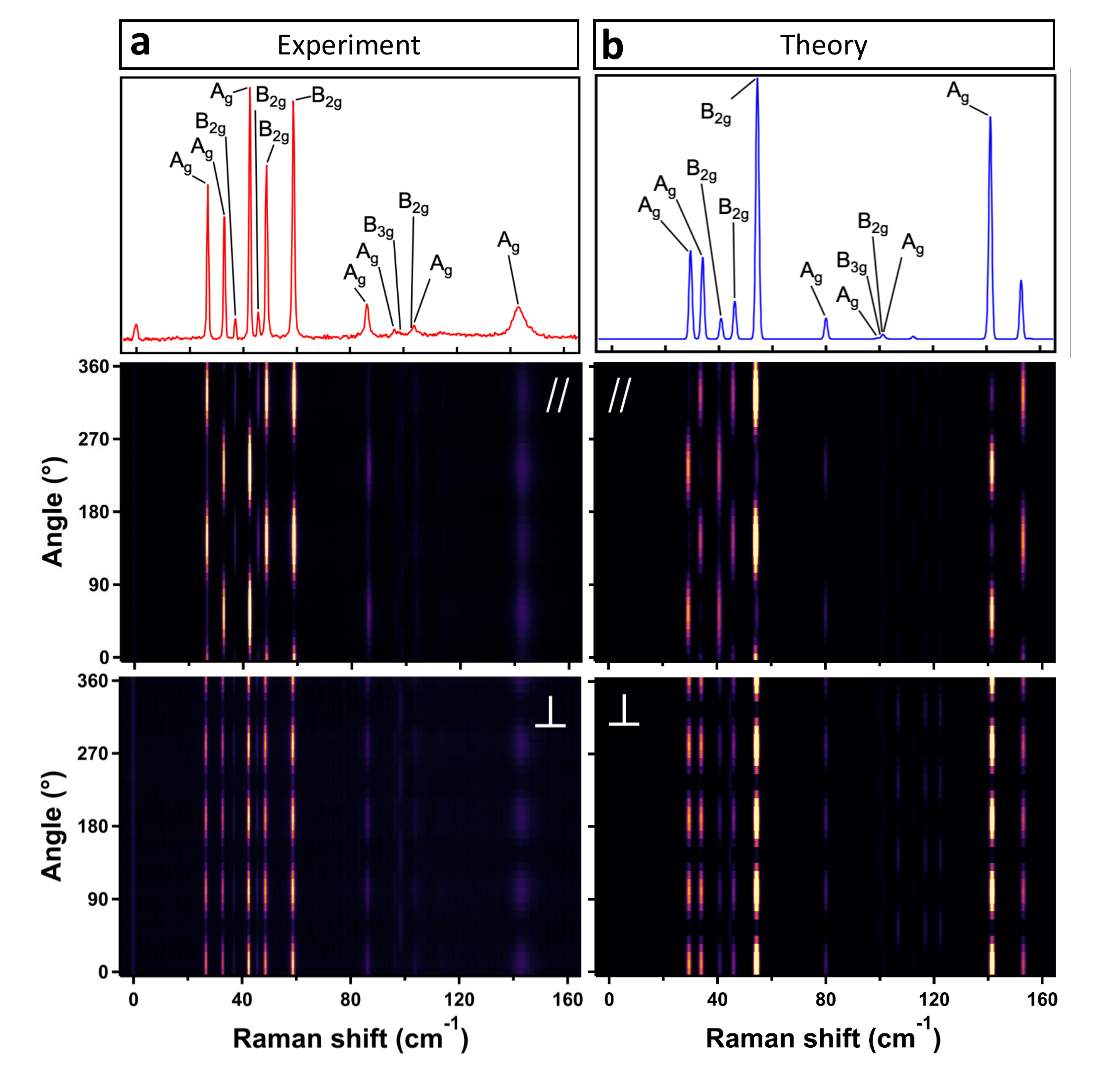}
		\caption{(a) Experimental and (b) Theoretically calculated angular dependence of polarized Raman spectra of the orthorhombic phase of MAPI at 10 K. The top panel shows the averaged spectra (over all angles), middle panel shows parallel and lower panel shows cross configuration for experiment and theory respectively. $\theta$ = 0$^\circ$ is an arbitrary angle and theory is rotated to align with experiment.}
		\label{PORaman}
	\end{figure*}
	
	Fig.~\ref{PORaman}a shows the contour plot of the Raman spectra of MAPI acquired at different excitation polarization angles. $\theta $ = 0$\mathrm{{}^\circ}$ is an arbitrary angle with respect to the sample axes, determined by the experimental setup. The upper panel shows the averaged Raman spectrum (over all angles), while the middle and lower panels show the spectra in parallel and cross configurations respectively. A periodic modulation in intensity as a function of polarization angle is apparent for all Raman-active modes. All the modes in parallel configuration show a periodicity of 180$^\circ$ except the mode at $\approx$98 cm$^{-1}$, which shows a 90$^\circ$ periodicity. In the cross configuration, all modes possess a 90$^\circ$ periodicity. It is worth mentioning that all the data presented here are in raw form without any normalization or base-line correction. The absence of any artifacts, intensity leakage or drift demonstrates the excellent stability of our Raman set-up and the absence of material evolution or damage during the experiment. Each Raman spectrum was deconvolved using a multi-Lorentz oscillator model (details in supplemental material (SM)\cite{SM}) and the intensity angular dependence was extracted.
	
	The quantitative analysis of the angular dependence of the scattered intensity using Equation~(\ref{Placzek}) requires prior knowledge of the crystal face probed in the PO Raman experiment. As MAPI changes phase to a lower symmetry (spontaneous symmetry breaking) as it is cooled, XRD taken at room-temperature doesn't provide the proper orientation of the crystal surface. Therefore, it is necessary to obtain exact information about the crystal orientation at cryogenic temperature using XRD at the same spot where Raman is probed. This is experimentally challenging. Therefore, we turned to DFPT calculations to determine the orientation of the crystal surface. 
	
	\subsection{Determination of illuminated face and mode assignments using DFPT}
	To determine the orientation of the illuminated face, we first used \textit{ab-initio} DFPT to calculate the unpolarized Raman spectrum in the low-frequency regime as shown in the top panel of Fig.~\ref{PORaman}b. Thereafter, the corresponding Raman tensors were derived through a procedure described in the SM\cite{SM}. Using the calculated Raman tensors, the angular dependence of the Raman intensities was simulated for different crystal orientations. The experimental angular dependence best matched with that calculated along the [101]  crystal orientation, as shown in Fig.~\ref{PORaman}b middle and lower panels.
	
	Once we determined the correct orientation, we proceeded to complete the identification of the mode symmetries and their dominant motions (Table \ref{Table1}) by matching calculated and experimental mode frequencies and polarization dependence. 
	
	\begin{table}[ht!]
		\caption{\ref{Table1} Frequency, symmetry, and corresponding dominant atomic motions in MAPI}	
		\begin{ruledtabular}	
			\begin{tabular}{ccccc}
				Mode&\multicolumn{2}{c}{Frequency (cm$ ^{-1} $)}&Symmetry	&Dominant Motion\\
				&Experiment&Theory&        &                        \\
				\hline
				1&	26.9&	29.4&	A$ _{g}$&	    PbI$ _{3}$ Network\\
				2&	33.1&	33.9&	A$ _{g}$&	     PbI$ _{3}$ Network\\
				3&	37.2&	35.6&	B$ _{2g}$&	PbI$ _{3}$ Network\\
				4&	42.6&	40.8&	A$ _{g}$&	    PbI$ _{3}$ Network\\
				5&	45.8&	---&	B$ _{2g}$&	---\\
				6&	48.8&	45.9&	B$ _{2g}$&	PbI$ _{3}$ Network\\
				7&	58.9&	54.2&	B$ _{2g}$&	PbI$ _{3}$ Network\\
				8&	86.5&	80.0&	A$ _{g}$&	    MA motion\\
				9&	96.7&	99.3&	A$ _{g}$&	    PbI$ _{3}$ Network\\
				10&	98.7&	100.0&	B$ _{3g}$&	PbI$ _{3}$ Network\\
				11&	102.7&	101.1&	B$ _{2g}$&	PbI$ _{3}$ Network\\
				12&	104.4&	101.2&	A$ _{g}$&	    PbI$ _{3}$ Network\\
				13&	143.3&	141.4&	A$ _{g}$&	    MA motion\\
				
			\end{tabular}
		\end{ruledtabular}
		\label{Table1}
	\end{table}
	
\begin{figure}[ht!]
		\includegraphics[width=3.5in]{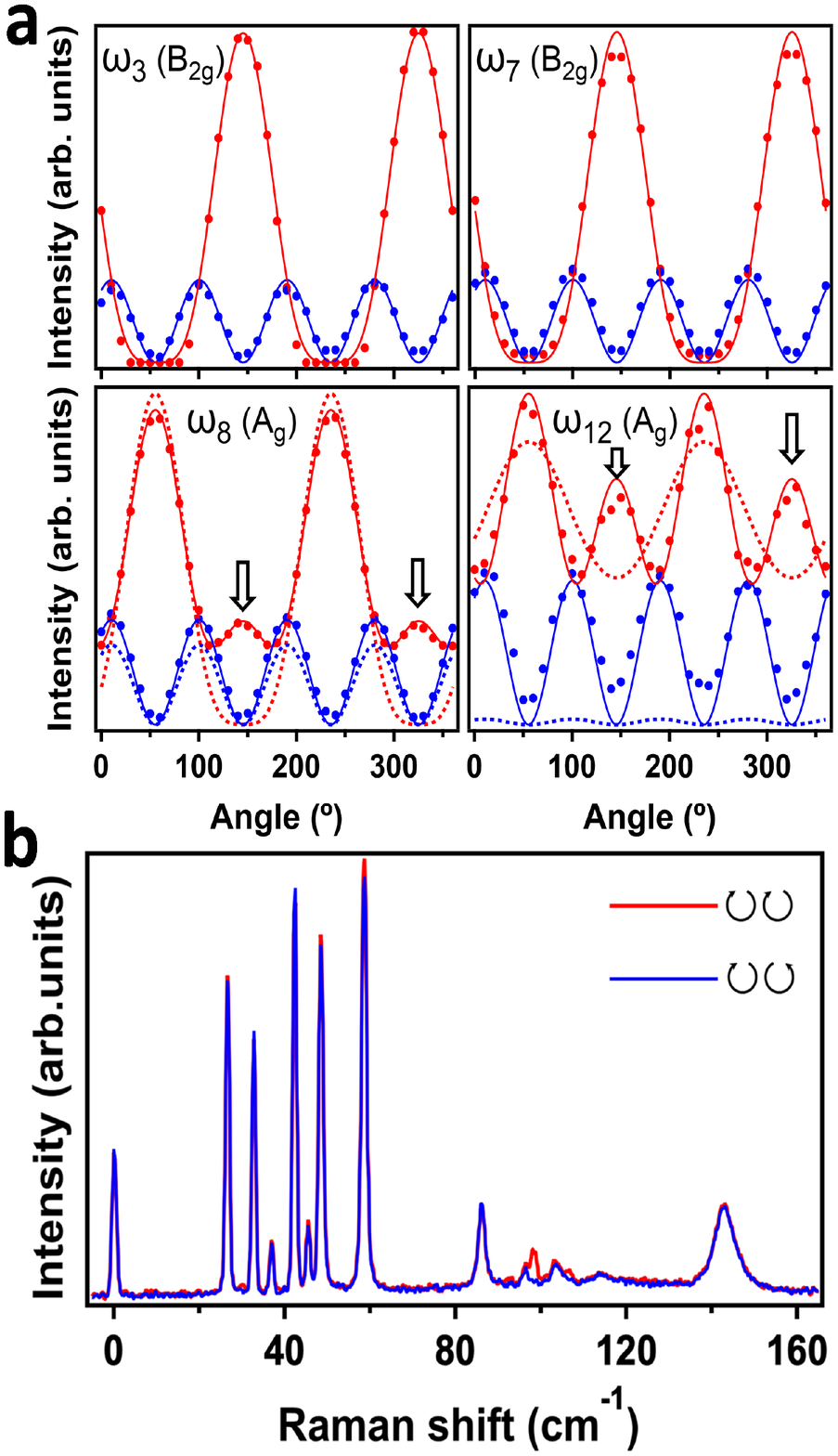}
		\caption{(a) Fitting of PO data using equation (\ref{ModifiedPlaczek}). The dotted curve is the fitting results using equation (\ref{Placzek}) without the consideration of optical birefringence. The secondary maximum for A$ _{g} $ modes is shown by the arrows. (b) Circular polarization Raman spectra of MAPI in parallel (co-rotating) and cross (counter-rotating) configurations.}
		\label{Ramanfitting}
	\end{figure}

	Having determined the correct Raman tensor for each mode, we noticed that the angular dependencies of the A${}_{g}$ modes could not be satisfactorily fit with Equation \ref{Placzek}, although the B${}_{2g}$ modes fit quite well. This is illustrated in Fig.~\ref{Ramanfitting}a where the experimental data are shown by dots and the best fit to Equation (\ref{Placzek}) is shown by dashed lines. Particularly, the secondary maximum of the A$ _{g} $ mode for the parallel polarization configuration was never reproduced by the fitting
	curves (labeled with arrow in Fig.\ref{Ramanfitting}a). We therefore conclude that the conventional
	Equation \ref{Placzek} fails to explain the behavior of the A$ _{g} $ Raman modes. To resolve this discrepancy, we now turn to circularly polarized Raman measurements.
	
	\subsection{Detection of birefringence effects using circularly polarized Raman scattering}
	Fig.\ref{Ramanfitting}b shows the Raman spectra acquired using circularly polarized light (CPL) excitation. In isotropic materials, a Raman mode would retain its light polarization even when excited with CPL (See SM for Silicon as example system). The scattered CPL can be analysed for the component rotating in the same (co-rotating,  $\circlearrowright \circlearrowright $) or opposite (contra-rotating, $ \circlearrowright \circlearrowleft $) direction with respect to the incident polarization. The reversal coefficient $\varrho $, along the [101] direction, is defined as\cite{Nestro1973}
	
	\begin{equation} 
	\varrho =\frac{I_{\circlearrowright \circlearrowleft }}{I_{\circlearrowright \circlearrowright }}=\frac{{\left|\left(a+c\right)+2b-2e\right|}^2}{{\left|\left(a+c\right)-2b+2e-i\sqrt{8}\left(d-f\right)\right|}^2} 
	\label{ReversalCoef}
	\end{equation} 
	
	\noindent where $I_{\circlearrowright \circlearrowleft }$ and  $I_{\circlearrowright \circlearrowright }$ are the scattering intensities of the counter-rotating and co-rotating components, respectively and a, b, c, d, e, f are the components of the Raman tensors defined in section \ref{PO}.  
	It is apparent that $\varrho =1$ for all observed modes, except the 98.7 cm$^{-1}$ mode which presents $\varrho =0$. The reversal coefficient 1 indicates that the polarization of the CPL is not maintained after scattering. We hypothesize that this is a manifestation of the birefringence effect arising from the anisotropic properties of HOPs\cite{Wang2016,Cho2016,Leguy2016b}. To test our hypothesis, we modify the Equation~(\ref{Placzek}) to accommodate birefringence effects:
	
	\begin{equation} 
	I\left(\theta \right)\propto \left|e_s(\theta +\delta)J^*{\mathcal{R}}^TR\mathcal{R}Je_i(\theta +\delta)\right|^2
	\label{ModifiedPlaczek} 
	\end{equation} 
	
	and
	
	\begin{equation} 
	J=\left( \begin{array}{ccc}
	1 & 0 & 0 \\ 
	0 & e^{i\left|{\phi }_y\right|} & 0 \\ 
	0 & 0 & e^{i\left|{\phi }_z\right|} \end{array}
	\right) 
	\end{equation}
	
	\noindent where \textit{J} is the Jones matrix of light propagation in the material \cite{Jones1941} and * denotes the complex conjugate transpose. $\left|{\phi }_y\right|=|{\phi }_{yy}-{\phi }_{xx}|$ and $\left|{\phi }_z\right|=|{\phi }_{zz}-{\phi }_{xx}|$ are the relative phases that the electric field components accumulate during propagation. Since the Raman measurements were done on a single crystallographic plane, $\left|{\phi }_z\right|$ cannot be determined by this specific model and cancels out during the calculation (see SM\cite{SM}). An additional phase factor $\delta$ is added to the angle $\theta $ to accomodate the birefringence effect so that ($\theta $+$\delta$) is the actual angle of light polarization with respect to the crystal's \textit{X}-axis. Since the wavelength of scattered light varies negligibly compared to the incident as regards birefringence, $\phi_y$ and $\delta$ were held constant across all modes. 
	
	As shown in Fig.~\ref{Ramanfitting}a, Equation (\ref{ModifiedPlaczek}) reproduces the intensity angular dependence of all the modes. Furthermore, we calculated the reversal coefficients (Equation (\ref{ReversalCoef})) using the obtained Raman tensor components as additional verification to the fit results and got an excellent agreement between the model and experimental results (See SM\cite{SM}, Table S2), including $\varrho$=0 for the mode at $\approx$98 cm$^{-1} $. Previous work by Osorio et al.\cite{Perez-Osorio2018} measured polarized Raman of MAPI at a single angle and analysed the symmetry using isotropically averaged Raman intensity. Consequently, their assignments slightly differ from ours. 
	
	\section{Conclusion}
	In conclusion, we have presented in detail a method to study the symmetries and atomic motions of the Raman-active phonons in HOPs, combining linearly and circularly polarized light based polarization-orientation Raman measurements and density functional perturbation theory. This approach allowed us to unambiguously assign all the low frequency lattice modes to specific symmetries. The anomalous behavior of the A$ _{g}$ modes is attributed to strong birefringence effects. Furthermore, we could calculate the crystal orientation, which is difficult to obtain by conventional diffraction methods at cryogenic temperatures. Thus, the combination of experimental polarization-dependent Raman with theoretical calculations can provide additional and more accurate information about the structural dynamics of the material, which otherwise was inaccessible or difficult to acquire.
	
	\begin{acknowledgments}
	
	\noindent The authors  would  like  to  thank Dr. Tsachi Livneh (NRC) for fruitful discussions,  Dr. Ishay Feldman (WIS) for performing X-Ray diffraction measurements and Dr. Lior Segev (WIS) for developing the Raman software. R. S. acknowledges FGS-WIS for financial support. O. Y. acknowledges funding from: Israel Science foundation (1861/17), Israel-US Bi National Foundation (grant No. 2016650 , European Research Counsel (850041 - ANHARMONIC),  Benoziyo Endowment Fund for the Advancement of Science, Ilse Katz Institute for Material Sciences and Magnetic Resonance Research, Henry Chanoch Krenter Institute for Biomedical Imaging and Genomics, Soref New Scientists Start up Fund, Carolito Stiftung, Abraham \& Sonia Rochlin Foundation, Erica A. Drake and Robert Drake and the European Research Council. Also, this research is made possible in part by the historic generosity of the Harold Perlman Family. Z. D. and L. G. acknowledge support from the US National Science Foundation, under grant DMR-1719353.  J. Z. acknowledges support from a VIEST Fellowship at the University of Pennsylvania. A. M. R. acknowledges support from the Office of Naval Research under Grant N00014-17-1-2574. The authors acknowledge computational support from the High-Performance Computing Modernization Office.
	\end{acknowledgments}
	

	%\includepdf[pages={1-13}]{SI}
%%\clearpage
%%\setcounter{page}{1}}}
\clearpage
\newpage
\widetext
\begin{center}
	\large{\textbf{Halide perovskites under polarized light: Vibrational symmetry analysis using polarized
			Raman}}
\end{center}

\begin{center}
	Rituraj Sharma$ ^{1*} $, Matan Menahem$ ^{1*} $, Zhenbang Dai$ ^{2} $, Lingyuan Gao$ ^{2} $, Thomas M. Brenner$ ^{1} $, Lena Yadgarov$ ^{1} $, Jiahao Zhang$ ^{2} $, Yevgeny Rakita$ ^{1} $, Roman Korobko$ ^{1} $, Iddo Pinkas$ ^{1} $, Andrew M. Rappe$ ^{2\dagger} $, Omer Yaffe$ ^{1\ddagger} $
\end{center}

\begin{center}
	\textit{$^1$Department of Materials and Interfaces, Weizmann Institute of Science, Rehovoth 76100, Israel\\
		$^2$Department of Chemistry, University of Pennsylvania, Philadelphia, Pennsylvania 19104, USA}
\end{center}

\section*{SUPPLEMENTAL MATERIAL}

\setcounter{figure}{0}
\setcounter{page}{1}
\setcounter{equation}{0}
\setcounter{table}{0}
\setcounter{subsection}{0}

%\makeatletter\renewcommand{\theequation}{S\arabic{equation}}
\renewcommand{\thefigure}{S\arabic{figure}}
\renewcommand{\thetable}{S\arabic{table}}
\renewcommand{\theequation}{S\arabic{equation}}
\renewcommand{\thesubsection}{\arabic{subsection}}
\renewcommand{\bibnumfmt}[1]{[S#1]}
\renewcommand{\citenumfont}[1]{S#1}
%%%%%%%%%% Prefix a "S" to all equations, figures, tables and reset the counter %%%%%%%%%%

\subsection{Crystal Synthesis}
\noindent High quality, methylammonium lead iodide (MAPI) single crystals were grown at room temperature using the antisolvent method as discussed elsewhere\cite{Rakita2017}. The sample’s surface orientation, at room temperature, was determined by X-ray diffraction (XRD) and the polarization-orientation (PO) Raman was performed on the same face for which XRD was carried out.

\begin{figure*}[ht!]
	\includegraphics[width=3.4in]{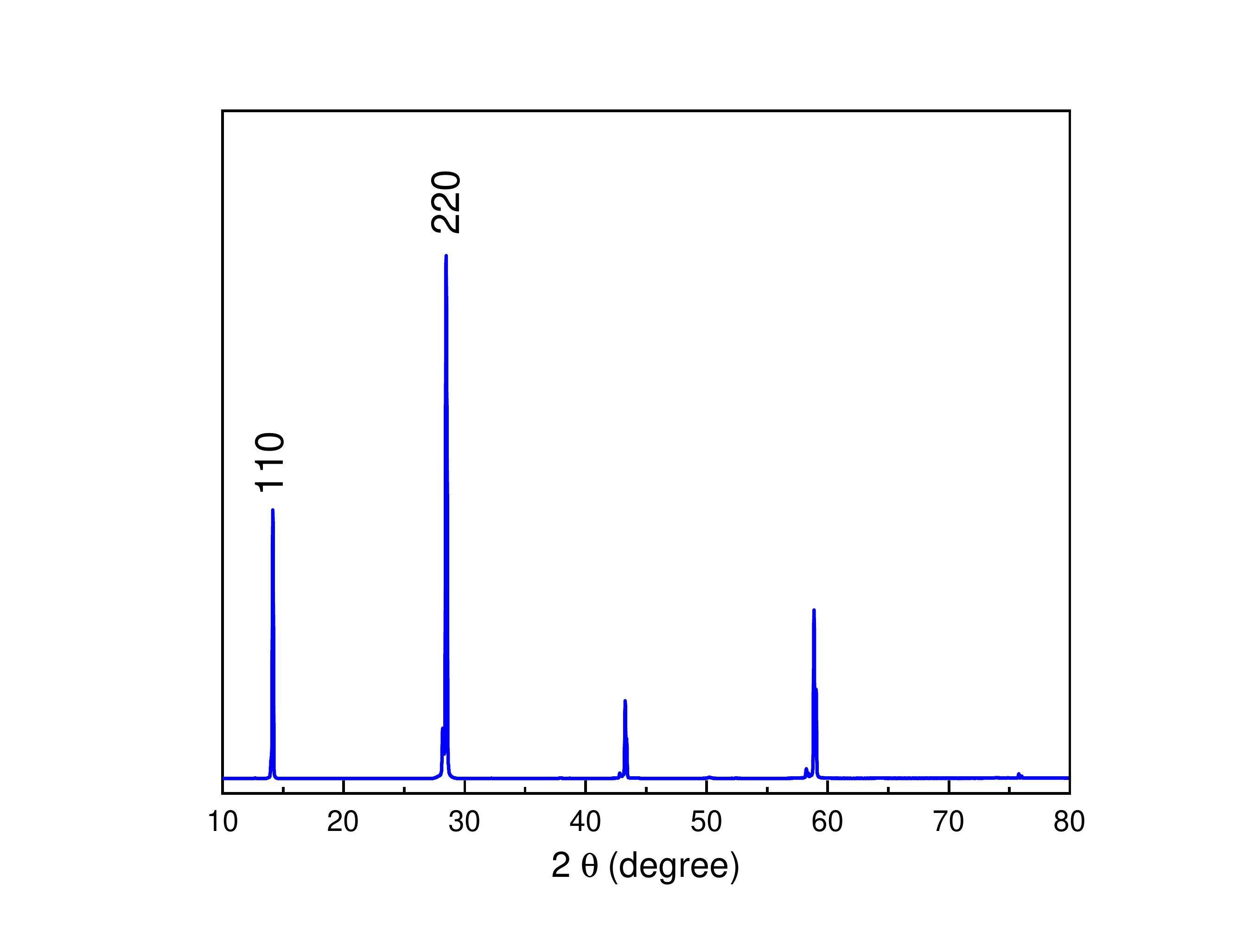}
	\caption{Room temperature XRD pattern of MAPI single crystal}
\end{figure*}

\subsection{Experimental set-up of PO Raman}

\noindent In our home built optical system, careful selection of optical components achieves high signal to noise ratio and extraordinary intensity stability with NIR excitations, under versatile conditions e.g. cryogenic to high temperatures, hydrostatic pressure and bias. Motorized optical components and integrated software further enables fast collection and analysis of data. 
In order to measure off-resonance PO Raman, MAPI sample was excited with 1.16 eV Nd:YAG laser (Coherent Inc., USA) with 2 mW excitation power, pre-treated with a set of two volume holographic grating (VHG) amplified spontaneous emission (ASE) filters (Ondax Inc., USA) and spatial filter equipped with 30 $\mu$m pinhole. The excitation beam was polarized using Glan-Laser polarizer and kept in the \textit{s}-polarization with respect to the mirrors in the system. The beam was then directed into a microscope (Zeiss, USA) and focused onto the sample via 0.42 NA / 50$ \times $ NIR objective (Zeiss, USA). The light polarization was controlled using a motorized zero-order half-wave plate (for linear polarization experiments) or quarter-wave plate (for circularly polarized experiments) located directly above the objective. The backscattered light was collected, linearized and filtered by another polarizer (analyzer). In order to detect the cross polarization, the beam was rotated by 90$ ^{\circ} $ using a motorized zero-order half-wave plate prior to the analyzer. In order to reduce Rayleigh scattering and increase signal to noise, the beam was passed through the VHG beam splitter and OD$ > $4 notch filters  (Ondax Inc., USA), and spatial filter equipped with 50 $\mu$m pinhole achieving spectral cutoff $ \pm $7 cm$ ^{-1}$ around the laser line. Finally, the beam was focused to 1 m long spectrometer (FHR 1000, Horiba) dispersed by 950 gr/mm IR blazed grating, achieving $\approx$0.3 cm$ ^{-1} $ spectral resolution, and detected by liquid N$ _{2} $ cooled InGaAs detector (Horiba Inc., USA). All optical components were purchased from Thorlabs Inc. unless otherwise mentioned. For high stability, the sample was mounted on motorized XY stage (Aerotech) with 100 nm drift stability. Furthermore, all motorized components were controlled and synchronized by LabView software. For cryogenic temperatures, the sample was mounted in liquid He cooled optical cryostat (Janis Inc. USA) under high vacuum 
(10 $ ^{-7} $ mbar at 10 K).

\subsection{Raman spectrum deconvolution of MAPI}

\noindent Following previously reported analysis\cite{Yaffe2017,Svitelskiy2003}, we analyzed the low frequency Raman modes of MAPI using multi damped Lorentz oscillator model. Each experimental spectrum of N Raman active modes was reconstructed using the following equation:

\begin{equation}
	I(\nu)=c_{BE}\sum_{i=1}^{N}c_i\frac{\left| \nu\right| \left| \nu_i\right| \Gamma_i^2}{\nu^2\Gamma_i^2+(\nu^2-\nu_i^2)^2}
\end{equation}

\noindent where $\nu$ is the spectral shift, $ \nu_i $ and $\Gamma_i$ are the resonance frequency and damping coefficient of the $ i $-th oscillator, $ c_i $ is a scaling factor to account for the mode’s intensity. 
Thermal phonon occupancy, described by the Bose-Einstein distribution, is accounted for in $  c_{BE} $ factor. Given an average population of $  n $, the Stokes scattering is proportional to $ n+1 $ while anti-Stokes scattering is proportional to $ n $

\begin{equation}
	c_{BE}(\nu) = \begin{cases}
		n(\nu)+1 & \quad \nu \ge  0, \\
		n(|\nu|) & \quad \nu < 0.
	\end{cases}
\end{equation}

\noindent where $ n(\nu)=\frac{1}{e^{h\nu/k_bT}-1} $, $h$ and $k_B$ being the Planck’s and Boltzmann’s constants respectively and T is the temperature.
We can calculate the crystal’s actual temperature by fitting the Stokes and anti-Stokes together, and indeed the temperature was 10 $\pm$ 1 K throughout the experiment. 
The temperature, oscillators’ frequency and damping should not depend at all on the excitation polarization. Therefore, in order to extract meaningful intensity angular dependence, all spectra were fitted while keeping these parameters constant.
Figure S1 shows the Raman spectra of MAPI at two angles of excitation, in the parallel configuration, along with the fit results and the individual peaks. The mode at 130 cm$ ^{-1} $ is artificial and is used to account for the background intensity around that frequency. The model could not properly capture the tail of the low-frequency peaks due to system’s response.

\begin{figure}[ht!]
	\includegraphics[width=3.4in]{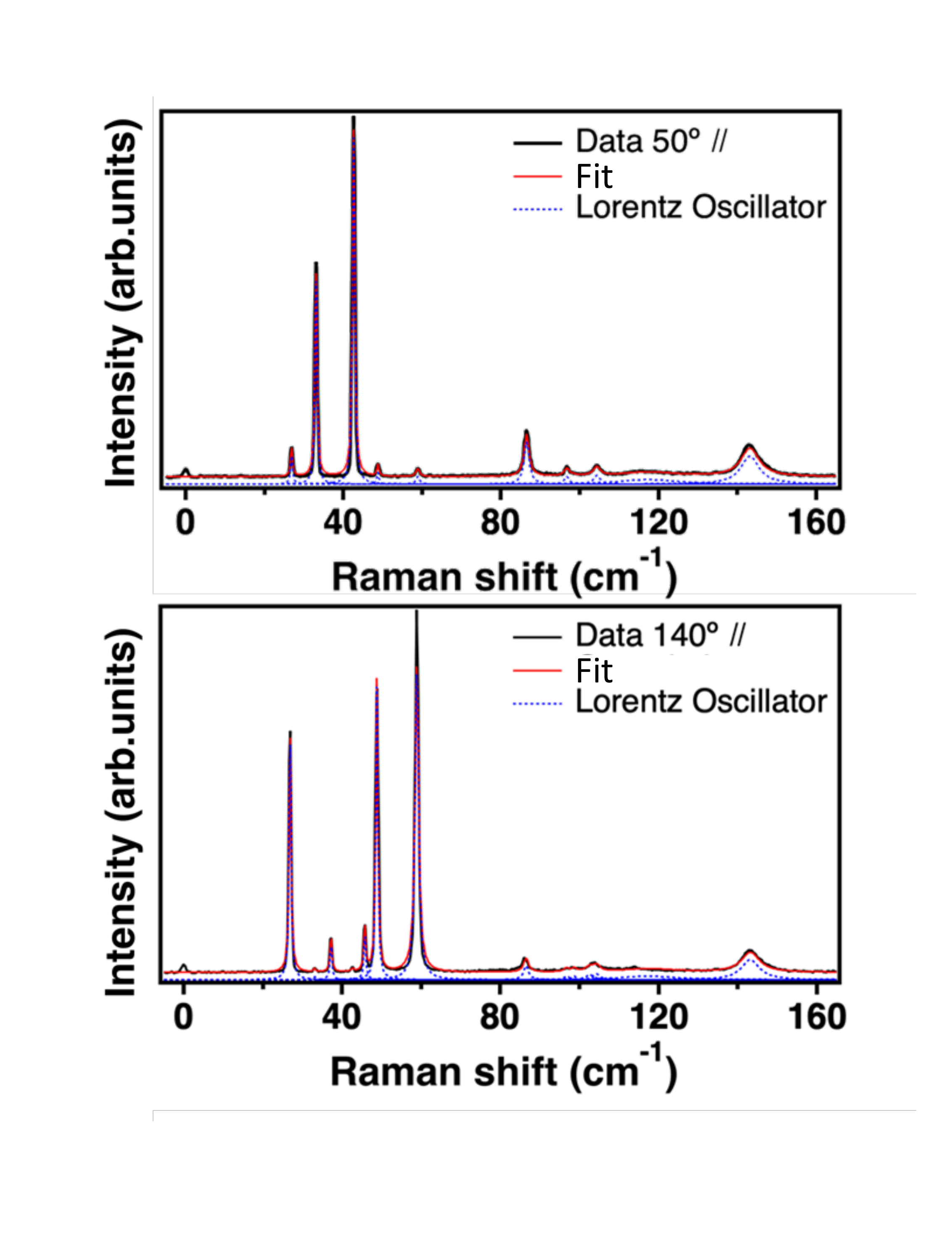}
	\caption{Polarized Raman spectra of MAPI at specific angle, the deconvolution fit result and the individual peaks.}
\end{figure}
\newpage
\subsection{Analysis of Raman scattering by circularly polarized light in Si}
\noindent In the classical description of Raman scattering, an optical mode should retain the light linearity (or circularity) during the scattering process. Therefore, the combination of linearly and circularly polarized light (LPL and CPL) has been used to derive the Raman tensor’s symmetry\cite{Lee2018,Liu1988,Thomsen1988}. In order to demonstrate the use of CPL, Si wafers of 3 different crystallographic orientations were excited with 1.16 eV CPL and the Raman scattering was measured (Figure S2).
Crystalline Si has cubic crystal structure corresponding to $ Fd-3m $ space group and the Raman active mode at 520 cm$ ^{-1} $ has the triply degenerate $ T_{2g} $ symmetry\cite{Steele2016}. The corresponding Raman tensors are:\\

\noindent $R_{T_{2g},xy}=\left( \begin{array}{ccc}
0 & d & 0 \\ 
d & 0 & 0 \\ 
0 & 0 & 0 \end{array}
\right)$, $R_{T_{2g},xz}=\left( \begin{array}{ccc}
0 & 0 & d \\ 
0 & 0 & 0 \\ 
d & 0 & 0 \end{array}
\right)$,  $R_{T_{2g},yz}=\left( \begin{array}{ccc}
0 & 0 & 0 \\ 
0 & 0 & d \\ 
0 & d & 0 \end{array}
\right)$. \\

\noindent The scattered intensity was calculated as:
\begin{equation}  
	I\left(\theta \right)\propto {\left|e_s(\theta )W^*{\mathcal{R}}^TR_{xy}\mathcal{R}We_i(\theta )\right|}^2+{\left|e_s(\theta )W^*{\mathcal{R}}^TR_{xy}\mathcal{R}We_i(\theta )\right|}^2+{\left|e_s(\theta )W^*{\mathcal{R}}^TR_{xy}\mathcal{R}We_i(\theta )\right|}^2
\end{equation} 
where $ e_i $ and $ e_s $ are the incident and scattered light polarization vectors:\\

\noindent $e_i=\left( \begin{array}{c}
cos(\theta) \\ 
sin(\theta) \\ 
0 \end{array}
\right)$, \noindent $e_{S,||}^T=\left( \begin{array}{c}
cos(\theta) \\ 
sin(\theta) \\ 
0 \end{array}
\right)$ and \noindent $e_{S,\perp}^T=\left( \begin{array}{c}
-sin(\theta) \\ 
cos(\theta)\\ 
0 \end{array}
\right)$;\\

\noindent $ \mathcal{R} $ is the rotation matrix between the laboratory system and the crystal orientation and $ W  $ is the Jones matrix of a retardation wave plate:

\begin{equation} 
	W=e^{i\sigma\pi}\left( \begin{array}{ccc}
		1 & 0 & 0 \\ 
		0 & -i & 0 \\ 
		0 & 0 & 0 \end{array}
	\right) 
\end{equation}

\noindent where $\sigma$ is the optical component’s retardance. CPL is achieved when $\sigma$ = 1/4 and $\theta$ = $\pi$/4. 
The reversal coefficient is defined as the ratio between the co-rotating and counter-rotating intensity. The reversal coefficient can be used to gain insight on the Raman tensor components or to verify the crystal orientation. Table S1 shows the great match between the calculated Si reversal coefficients from the model and the Raman spectra.

\begin{figure*}[ht!]
	\includegraphics[width=6in]{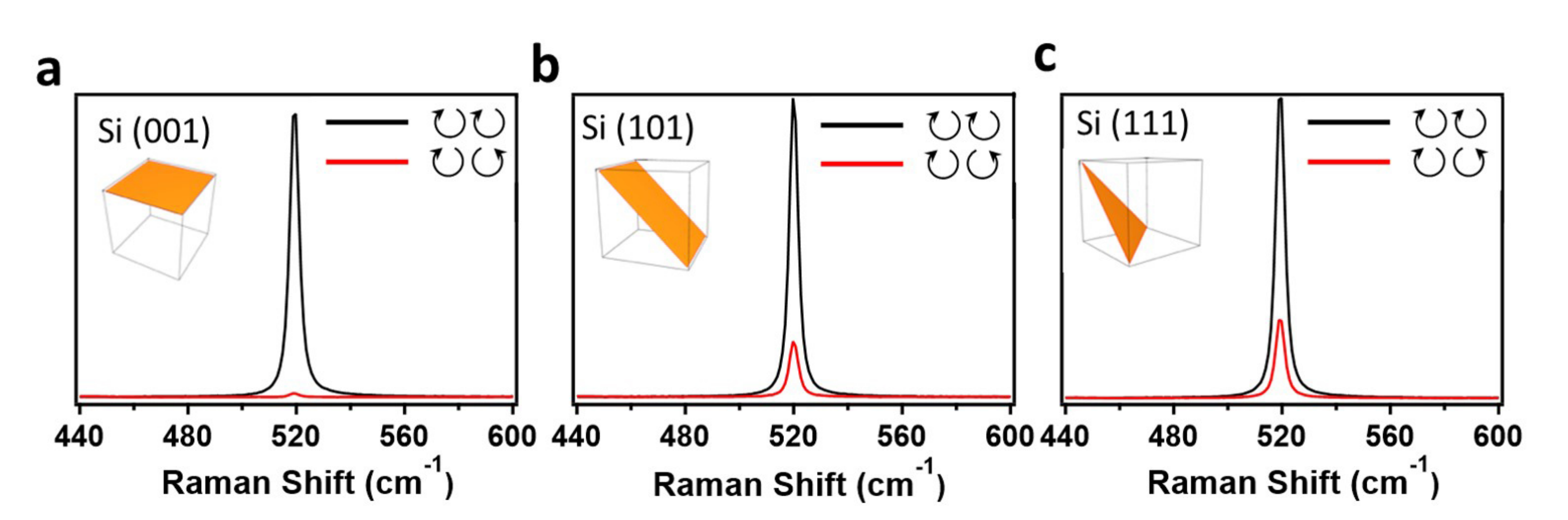}
	\caption{Raman scattering of c-Si wafer in the [001] (a) [101] (b) and [111] (c) crystallographic directions, excited with circularly polarized light. The black curves correspond to the co-rotating and the red curves correspond to the counter-rotating scattered light. Inset: schematic of the crystal plane in the cubic unit cell.}
\end{figure*}

\begin{table}[ht!]
	\caption{Calculated reversal coefficient of the Si Raman active mode, from experiment and model.}
	\begin{tabular}{ccc}
		\hline
		Crystal Orientation&	Experiment&	Model\\ 
		\hline
		001&0.01&0\\
		011&0.19&0.2\\
		111&0.26&0.25\\
		\hline
		
	\end{tabular}
\end{table}
\newpage
\subsection{Analysis of PO Raman spectra of MAPI} 
\noindent The Raman tensors corresponding to $ Pnma $ space group are:\\

\noindent $A_g=\left( \begin{array}{ccc}
a & 0 & 0 \\ 
0 & b & 0 \\ 
0 & 0 & c \end{array}
\right)$, $B_{1g}=\left( \begin{array}{ccc}
0 & d & 0 \\ 
d & 0 & 0 \\ 
0 & 0 & 0 \end{array}
\right)$,  $B_{2g}=\left( \begin{array}{ccc}
0 & 0 & e \\ 
0 & 0 & 0 \\ 
e & 0 & 0 \end{array}
\right)$ and $B_{3g}=\left( \begin{array}{ccc}
0 & 0 & 0 \\ 
0 & 0 & f \\ 
0 & f & 0 \end{array}
\right)$.\\

The modified intensity dependence is given as:
\begin{equation}  
	I\left(\theta \right)\propto {\left|e_s(\theta )W^*J^*{\mathcal{R}}^TR\mathcal{R}JWe_i(\theta )\right|}^2
\end{equation}

where $ e_i $, $  e_s $ and $ W $($\sigma$=0.5) are similar to the ones used for c-Si in section 4, and\\

$J=\left( \begin{array}{ccc}
1 & 0 & 0 \\ 
0 & e^{i\left |\phi_y\right |} & 0 \\ 
0 & 0 & e^{i\left |\phi_z\right |} \end{array}\right)$\\

In the laboratory system the light polarization is in the XY plane, while the crystal was measured perpendicular to [101] crystallographic direction. Therefore, the appropriate rotation matrix is\\

$ \mathcal{R}_{001\rightarrow101}=\frac{1}{\sqrt{2}}\left( \begin{array}{ccc}
1 & 0 & 1 \\ 
0 & \sqrt{2} & 0 \\ 
-1 & 0 & 1 \end{array}\right)$\\

\noindent The resulting intensity angular dependencies are:

\begin{equation}
	I_{||}(\theta) = \frac{1}{2}\left|(S-2e)cos^2(\theta)+be^{2i\phi_y}sin^2(\theta)+\frac{e^{i\phi_y}}{\sqrt{2}}(d-f)sin(2\theta)\right|^2
\end{equation}

\begin{equation}
	I_{\perp}(\theta) = \left|\frac{e^{i\phi_y}}{\sqrt{2}}(d-f)cos(2\theta)-\dfrac{1}{4}(S-2be^{2i\phi_y}-2e)sin(2\theta)\right|^2
\end{equation}

\noindent And the corresponding circularly polarized light intensities are
\begin{equation}
	I_{\circlearrowright\circlearrowright}(\theta) = \left|\frac{1}{2}(S-2e)cos^2(\theta)-be^{2i\phi_y}sin^2(\theta)-\frac{ie^{i\phi_y}}{\sqrt{2}}(d-f)sin(2\theta)\right|^2
\end{equation}

\begin{equation}
	I_{\circlearrowright\circlearrowleft}(\theta) = \left|\frac{ie^{i\phi_y}}{\sqrt{2}}(d-f)cos(2\theta)+\frac{1}{4}(S+2be^{2i\phi_y}-2e)sin(2\theta)\right|^2
\end{equation}

\noindent where $ S=(a+c) $ which could not be individually resolved with the current model.\\

\noindent The reversal coefficient is defined as:
$\varrho=\frac{I_{\circlearrowright\circlearrowleft}}{I_{\circlearrowright\circlearrowright}}=\frac{I_\perp(\theta=\frac{\pi}{4})}{I_{||}(\theta=\frac{\pi}{4})}=\begin{cases}
\frac{\left|S+2be^{2i\phi_y} \right|^2 }{\left|S-2be^{2i\phi_y} \right|^2} & \quad A_g \\
0 & \quad B_{1g} ~$ or $~ B_{3g} \\
1 & \quad B_{2g}	
\end{cases}$\\

Figure S4 shows the angular dependence as well as the fit to the model. For the mode at 98.7 cm$ ^{-1} $, the model cannot accurately reproduce the angular dependence, even though the assigned $ B_{3g} $ symmetry still provides the best match. This discrepancy between the model and data can be related to the difficulty to resolve the peak from the background in the PO Raman data.
Table S2 presents the resulting Raman tensor parameters. Because of the proportionality in the equation S5, only the ratio between Raman tensor components hold meaning and the values were normalized to highest Raman tensor component in the analysis. Another limit of the current model is that the sign of the tensor components cannot be resolved. Combining the result of the model with the theoretical Raman tensors, the sign of the individual tensor components as well as the ratio between $ a $ and $ c $ could be deduced.
\begin{figure*}
	\includegraphics[width=6in]{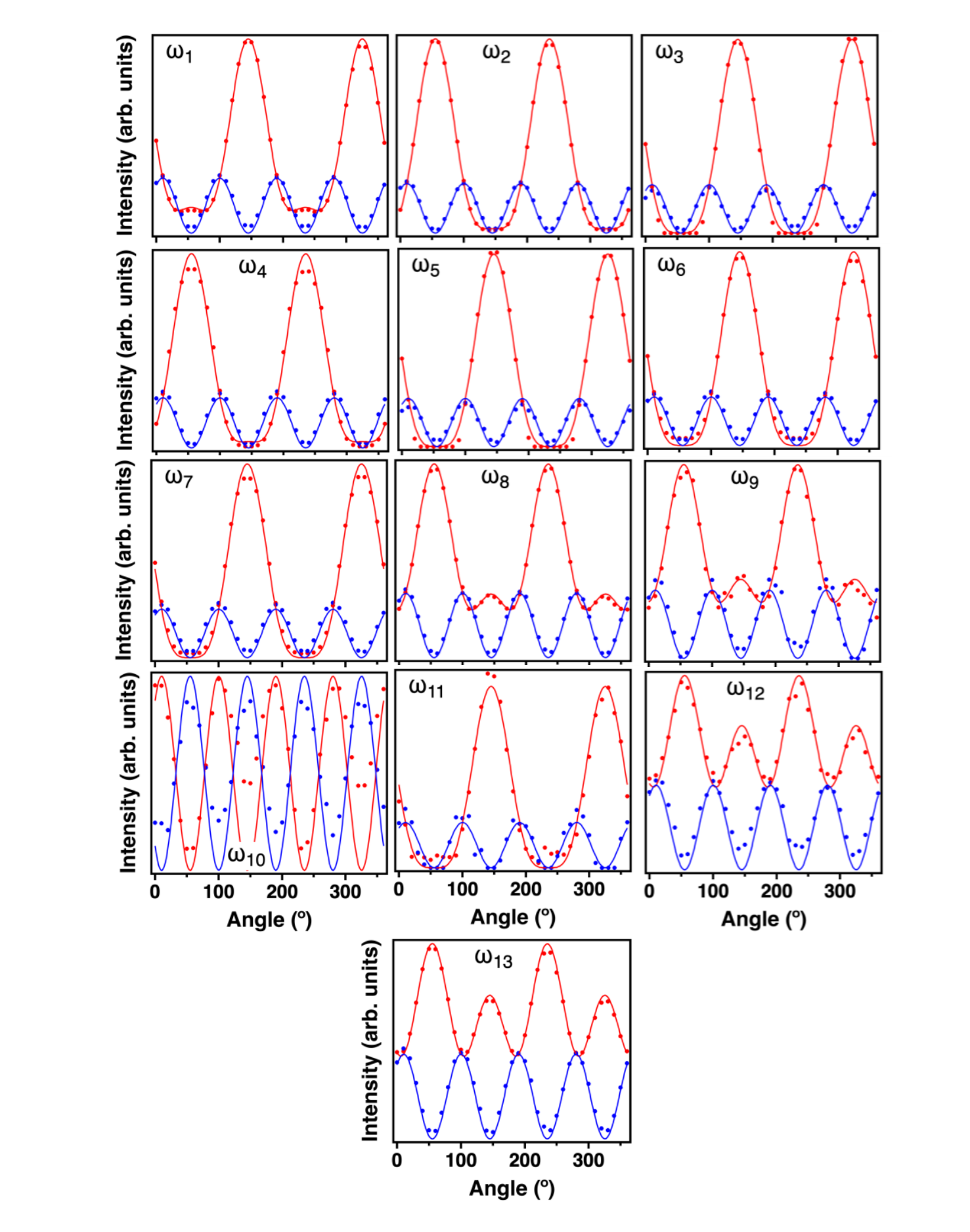}
	\caption{Intensity angular dependence and best fit to the model. Red curves correspond to the parallel configuration while blue curves correspond to the cross configuration.}
	
\end{figure*}
\newpage
\begin{table}
	\caption{Raman tensor components of the analyzed modes in MAPI.}
	\begin{tabular}{cccccccccll}
		\hline
		Mode Index&	Raman Shift (cm$ ^{-1} $)&	Symmetry&	a&	b&	c&	d&	e&	f& $ \varrho $ (exp)&$ \varrho $ (model)\\
		
		\hline
		1&	26.9&	A$ _{g} $&	0.583&	0.227&	0.668&	0&	0&	0&1 &1 \\
		2&	33.1&	A$ _{g} $&	-0.099&	0.675&	-0.096&	0&	0&	0&1.04 &1 \\
		3&	37.2&	B$ _{2g} $&	0&	0&	0&	0&	0.278&	0&0.90 & 1\\
		4&	42.6&	A$ _{g} $&	0.202&	0.940&	0.138&	0&	0&	0&1.05 &1 \\
		5&	45.8&	B$ _{2g} $&	0&	0&	0&	0&	0.351&	0&0.90 &1 \\
		6&	48.8&	B$ _{2g} $&	0&	0&	0&	0&	1&	0&0.98 & 1\\
		7&	58.9&	B$ _{2g} $&	0&	0&	0&	0&	-0.840&	0&0.97 &1 \\
		8&	86.5&	A$ _{g} $&	0.26&	0.306&	0.091&	0&	0&	0&1.03 &1 \\
		9&	96.7&	A$ _{g} $&	-0.100&	0.139&	0.077&	0&	0&	0&0.95 &1 \\
		10&	98.7&	B$ _{3g} $&	0&	0&	0&	0&	0&	0.193&0.1 &0 \\
		11&	102.7&	B$ _{2g} $&	0&	0&	0&	0&	-0.087&	0&0.91 & 1\\
		12&	104.4&	A$ _{g} $&	-0.108&	0.130&	0.116&	0&	0&	0&0.93 &1 \\
		13&	143.3&	A$ _{g} $&	-0.180&	0.168&	0.107&	0&	0&	0&0.98 &1 \\
		\hline& & 
	\end{tabular}
\end{table}

\subsection{Computation methods}

\noindent The nonresonant Raman tensor $ R^\nu $ can be evaluated under the harmonic approximation\cite{Bruesch1986g}:

\begin{equation}
	R_{ij}^\nu=\sum_{k\alpha}\frac{\partial^3\epsilon^{el}u_{k\alpha}^\nu}{\partial E_i\partial E_j\partial x_{k\alpha}\sqrt{M_k}}
\end{equation}

\noindent where $ \epsilon^{el} $ is the electronic energy of the system,  $ E_i $ is the $ i $-th component of a uniform electric field,  $x_{k\alpha}$ is the displacement of the $ k $-th atom in $ \alpha $ direction, $ u_{k\alpha}^\nu $ is one component of the eigenvector of the normal mode $\nu$, and $ M_k $ is the mass of the $ k $-th atom. Even though the dynamical correlation method can go beyond the harmonic approximation, for the current work we only focus on the low-temperature orthorhombic phase, and it suffices to adopt the harmonic approximation and use the more computationally efficient perturbation treatment.
Here, we use the Density Functional Perturbation Theory (DFPT) to evaluate the third-order derivative\cite{Lazzeri2003}. The initial structure is a $ \sqrt{2}\times2\times\sqrt{2} $ supercell obtained by relaxing the experimental-resolved structure under fixed lattice parameters (8.58 \AA , 12.63 \AA , 8.87 \AA)\cite{Weller2015}, and the calculation is performed via Density Functional Theory (DFT) with a $ 6\times4\times6 $ $ k $-mesh. The norm-conserving pseudopotential is used with local density approximation (LDA) for the exchange-correlation function. Grimme-d2\cite{Grimme2006} correction is employed for van der Waals interaction.  The criterion of structural relaxation is $\approx$0.005 meV$ \cdot $$ \AA^{-1} $, and this is good enough for the DFPT calculation. Acoustic sum rule (ASR) is applied to further eliminate the numerical error which breaks the translational symmetry. For periodic system, since Raman spectroscopy can only detect phonon modes at Gamma point, only one q point is needed in the DFPT calculation. Both the DFT and DFPT calculation are carried out in Quantum-Espresso\cite{Giannozzi2009}. 
Once the third-order derivative is obtained, we can construct the Raman tensor according to equation (S10).

\begin{figure*}
	\includegraphics[width=3.4in]{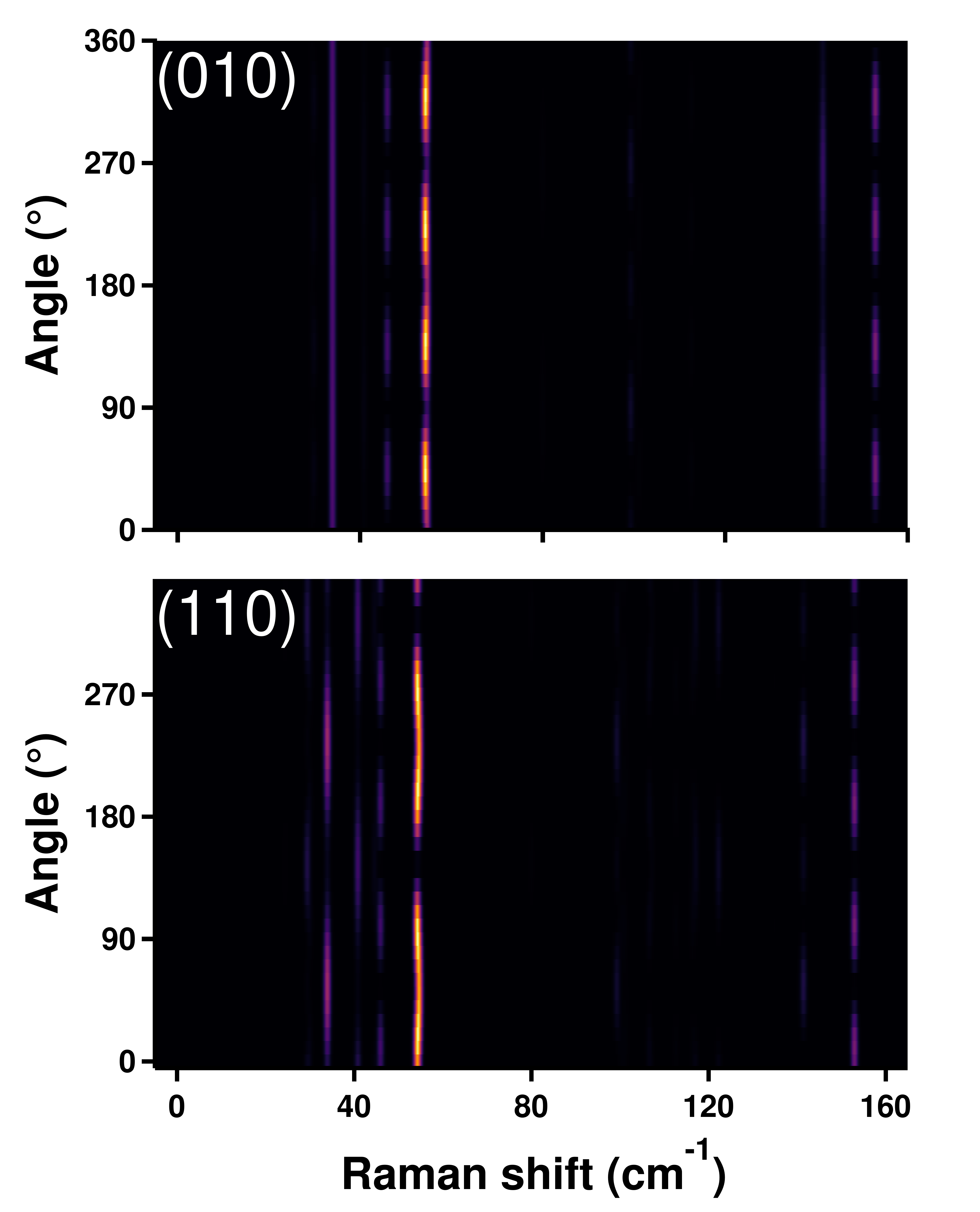}
	\caption{Theoretically calculated angular dependence of Raman spectra of MAPI for 010 and 110 crystal orientations.}
	
\end{figure*}
\newpage

\end{document}